\documentclass[10pt,aps,pra,twocolumn,showpacs,superscriptaddress,floatfix]{revtex4}
\usepackage{amsmath,amsfonts,algorithm,algorithmic}
\usepackage{bm}
\usepackage[pdftex]{graphicx}
  \graphicspath{{figures/pdf/}}
  \DeclareGraphicsExtensions{.pdf}
\def\vec#1{{\bf#1}}
\def\gvec#1{{\bm#1}}
\def\ket#1{| #1 \rangle}
\def\ip#1#2{\langle #1| #2 \rangle}
\def\bra#1{\langle #1 |}
\def\ave#1{\langle #1 \rangle}
\def\norm#1{\| #1 \|}
\def\Tr{\operatorname{Tr}}
\def\diag{\operatorname{diag}}
\def\D{\mathcal{D}}
\def\ONE{\bm 1}

\begin{document}
\title{Quantum System Identification by Bayesian Analysis of Noisy Data:
Beyond Hamiltonian Tomography}

\author{S.~G.~Schirmer}\email{sgs29@cam.ac.uk} 
\affiliation{Department of Applied Maths and Theoretical Physics,
University of Cambridge, Cambridge CB3 0WA, United Kingdom}

\author{D.~K.~L.~Oi}\email{daniel.oi@strath.ac.uk} 
\affiliation{SUPA, Department of Physics, University of Strathclyde,
Glasgow G4 0NG, United Kingdom} \date{\today}

\begin{abstract}
  We consider how to characterize the dynamics of a quantum system from
  a restricted set of initial states and measurements using Bayesian
  analysis. Previous work has shown that Hamiltonian systems can be well
  estimated from analysis of noisy data.  Here we show how to generalize
  this approach to systems with moderate dephasing in the eigenbasis of
  the Hamiltonian.  We illustrate the process for a range of three-level
  quantum systems.  The results suggest that the Bayesian estimation of
  the frequencies and dephasing rates is generally highly accurate and
  the main source of errors are errors in the reconstructed Hamiltonian 
  basis.
\end{abstract}

\maketitle

\section{Introduction}

Recent advances in nanofabrication technology increasingly enable the
construction of devices operating in the quantum regime.  However, to
utilize coherence effects for practical applications such as quantum
information processing and communication tasks requires the ability to
engineer their dynamics with high precision.  Considerable progress in
the area of laser technology and optimal control has shown that precise
coherent manipulation of the dynamics is not infeasible for a variety of
quantum systems, but such control requires accurate knowledge of the
system's dynamical behavior and response to external fields, which can
be used to construct accurate models from which effective control
designs can be engineered.  The problem is particularly acute for
manufactured systems, due to inevitable variations in the manufacturing
processes, which ensure that the exact behavior of each device is unique
and must be individually measured and characterized.

For the manufacture of large-scale practical devices, the design and operation
of each device should be as simple as possible meaning that the physical
resources available to initialize and measure the state of a system are usually
restricted to a single basis set defined by static electrode geometry.  In
normal operation, any state can be produced from an initial fiducial one by
applying a suitable unitary rotation.  This also enables us to effectively
perform measurements in an arbitrary basis, and given both these abilities, one
can perform quantum process tomography~\cite{qpt1,qpt2}.  However, the problem
of characterizing a device is not trivial since initially, if one does not yet
know the control response of the system, one cannot generate the unitary
rotations required in the first place, leading to a Catch-22 situation.

What is required is a method of bootstrapping the control and
characterization process so that the system dynamics and response can be
incrementally assessed until full control and process tomography is
possible, and only using the \emph{in situ} resources.  Hence, we have
developed techniques based upon the analysis of generalized coherent
oscillation data from Rabi or Ramsey-type experiments. There are several
approaches to the analysis of such experimental data including
frequency-domain and time-domain analysis.  In the regime of a single
system transition, Fourier analysis is effective but in the presence of
multiple signals, it ceases to an optimal estimator.

In previous work, we have shown how Bayesian signal analysis can be
effective in determining accurate model parameters in generic two-qubit
Hamiltonian systems where multiple frequencies are present.  In this
work, we extend the technique to systems with dephasing and use Bayesian
signal analysis to reconstruct the underlying dynamics, which are now
non-unitary.  We apply this technique to three-level (qutrit) systems
and analyze its performance for a range of dephasing rates and find that
as long as coherent dynamics dominate, which would be the case for
quantum information purposes, signal parameters can generally be
reliably extracted and the system effectively reconstructed.

\section{Open quantum systems}

The evolution of a closed quantum system is governed by a time-dependent
unitary operator $U(t)$ obeying the Schrodinger equation.  The evolution 
of an open quantum system can be highly complicated but under certain 
conditions it can be described by a master equation
\begin{equation} 
\label{eq:LME}
 \dot{\rho}(t) = -i[H,\rho(t)]+L_D\rho(t),
\end{equation}
where $[A,B]=AB-BA$ is the usual matrix commutator, $L_D$ is a
super-operator describing the interaction with the environment, and
$\rho$ is a unit-trace positive operator $\rho$ on $H$ representing the
state of the system.  This form of master equation is generally
applicable to systems interacting with a memory-less (Markovian)
reservoir such as an effectively infinite bath, where it can be shown
that the superoperator $L_D(\rho)$ takes the form $L_D(\rho) = \sum_k
\D[V_k]\rho$, where $V_k$ are operators on $H$ and the superoperators
$\D[V_k]$ are defined by
\begin{equation}
  \D[V_k]\rho = V_k \rho V_k^\dag 
                -\frac{1}{2}(V_k^\dag V_k\rho + \rho V_k^\dag V_k).
\end{equation}
Under certain conditions we can make further simplifying assumptions.
For example, dissipative effects in open systems weakly coupled to an
environment are often dominated by a certain types of decoherence such
as pure phase relaxation or population relaxation processes such as the
spontaneous emission of photons or phonons.  These types of processes
can be described by relatively simple master equations.  In the case of
pure dephasing the dissipation superoperator is often determined by a
single Hermitian operator $V$.  In this case, it is easy to show that
the master equation simplifies
\begin{equation}
  \label{eq:LME2}
  \dot\rho(t) = -i [H,\rho(t)] - \frac{1}{2} [V,[V,\rho(t)]].
\end{equation}
Even with these simplifying assumptions on the open system dynamics we
see that full system identification now requires the identification two
generally independent Hermitian operators $H$ and $V$, which in general
means the identification of $2(N^2-1)$ real parameters. Fortunately,
dephasing often acts in the eigenbasis of the Hamiltonian, in which case
$H$ and $V$ commute and are simultaneously diagonalizable, i.e., there
exists a basis $\{\ket{e_\nu}\}$ such that
\begin{equation}
  H = \sum_{\nu=1}^N \lambda_\nu \ket{e_\nu}\bra{e_\nu}, \quad
  V = \sum_{\nu=1}^N \gamma_\nu \ket{e_\nu}\bra{e_\nu}
\end{equation}
where $\lambda_\nu$ and $\gamma_\nu$ are real, and in this case the
identification problem reduces to finding a joint eigenbasis
$\{\ket{e_\nu}\}$ and the corresponding eigenvalues $\lambda_\nu$ and
$\gamma_\nu$ of $H$ and $V$, respectively.  This simplifies the problem.
If $\tilde{H}=\diag(\lambda_\nu)$, $\tilde{V}=\diag(\gamma_\nu)$ and
$\tilde{\rho}$ is the representation of the state in a joint eigenbasis
of $H$ and $V$ then it is easy to see that the master equation
(\ref{eq:LME}) gives
\begin{equation}
  \frac{d}{dt}\tilde{\rho}_{\mu\nu} (t) = 
  -i(\omega_{\mu\nu}-i\Gamma_{\mu\nu})\tilde{\rho}_{\mu\nu}
\end{equation} 
where $\omega_{\mu\nu}=\lambda_\mu-\lambda_\nu$ and 
$\Gamma_{\mu\nu}=\frac{1}{2}(\gamma_\mu-\gamma_\nu)^2$, i.e., we have
\begin{equation}
  \tilde{\rho}_{\mu\nu}(t) 
  = e^{-it(\omega_{\mu\nu}-i\Gamma_{\mu\nu})} \tilde{\rho}_{\mu\nu}(0).
\end{equation}
and if $W$ is the unitary basis transformation that maps the measurement
basis to the joint eigenbasis of $H$ and $V$, then the evolution of the
density operator $\rho$ with respect to the measurement basis is given
by $\rho(t)=W^\dag\tilde{\rho}(t) W$.  Thus the evolution is determined
by the transition frequencies $\omega_{\mu\nu}$, dephasing rates
$\Gamma_{\mu\nu}$ and the relation between the system and measurement
basis $W$, which are to be determined.  

\section{Experimental Identification Protocol}
\label{sect:exp}

As in previous work~\cite{SKO2004,SKOC2004,CSGWOH2005,CGOSWH2006,
DSOCH2007,BIRS2007,SO2009} we assume that we can prepare and measure the
system in a fixed set of (orthonormal) computational basis states
$\{\ket{1},\ket{2},\ldots,\ket{N}\}$, where $N$ is the Hilbert space
dimension.  No other measurements or resources such as non-basis states
are assumed to be available initially.  The basic protocol is to prepare
the system in a computational state, let it evolve for a period of time,
then measure the probabilities that the system ends up in one of the
computational basis states, repeating it for different times and all
computational basis states.  The experimental data thus consists of
$N^2$ time traces, $p_{\ell,k}(t)$, with $k,\ell=1,2,3$, which
represents the probability that the system, initially in state $\ket{k}$
is measured in state $\ket{\ell}$ after evolving under the system
Hamiltonian for time $t$.

\begin{figure}
\includegraphics[width=0.5\textwidth]{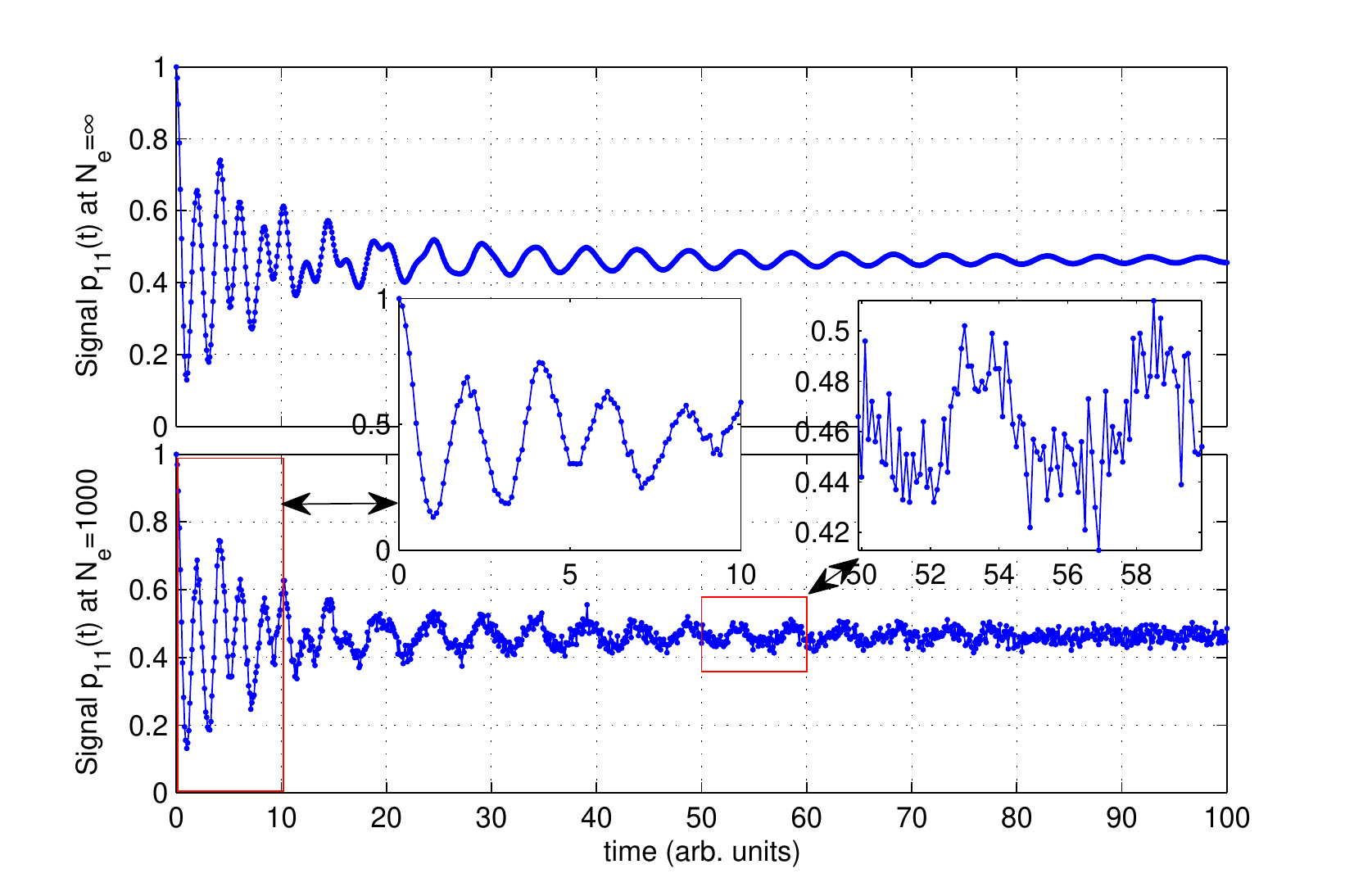}
\caption{Simulated measurement traces of a Qutrit. Ideal signal trace with no
  projection noise (top) and with 1000 repetition samples per time point
  (bottom). The signal consists of three damped sinusoids and represents the
  probability of measuring the system to be in the computational state $\ket{0}$
  if it was original initialized in $\ket{0}$. At long times, noise dominates
  the signal which leads to an optimal total sampling time.}
\label{fig:DataSignals}
\end{figure}

When we include dephasing in the Hamiltonian eigenbasis, it can be shown
that the observable probabilities are
\begin{equation}
\label{eq:probs}
\begin{split}
 p_{k\ell}(t) 
 =& c_{k\ell}(t) + 2\sum_{\nu>\mu} 
      a_{k\ell;\mu\nu} e^{-t\Gamma_{\mu\nu}}\cos(\omega_{\mu\nu}t)\\ 
  & \qquad\qquad\quad
    + b_{k\ell;\mu\nu} e^{-i\Gamma_{\mu\nu}}\sin(\omega_{\mu\nu}t),
\end{split}
\end{equation}
where the coefficients are
\begin{subequations}
\label{eq:coeff}
\begin{align}
  a_{k\ell;\mu\nu} &= s_{k\ell;\nu} s_{k\ell;\mu} \cos(\Delta_{k\ell;\mu\nu}), \\
  b_{k\ell;\mu\nu} &= s_{k\ell;\nu} s_{k\ell;\nu} \sin(\Delta_{k\ell;\mu\nu}), \\
  c_{k\ell}        &= \textstyle\sum_\nu s_{k\ell;\nu}^2.
\end{align}
\end{subequations}
Here $s_{k\ell;\nu}$ and $\delta_{k\ell;\nu}$ are the amplitude and
phase of the complex number $\ip{\ell}{\xi_\nu}\ip{\xi_\nu}{k}$ and
$\Delta_{k\ell;\mu\nu}=\delta_{k\ell;\nu}-\delta_{k\ell;\mu}$ is the
phase difference.  

If the Hamiltonian is known to be real-symmetric in the computational
basis, which is the case for many systems including atomic and molecular
systems, where the off-diagonal elements of the Hamiltonian are usually
real transition strengths or dipole moments, and spin systems, the
problem can be simplified.  The eigenvectors of a real-symmetric matrix
are real, thus the phases $\delta_{k\ell;\nu}$ must be multiples of
$\pi$ so that $e^{i\delta_{k\ell;\nu}} =\pm 1$, and since the sine of a
multiple of $\pi$ vanishes, we have $b_{k\ell;\mu\nu}=0$. In many cases
the signs of the off-diagonal matrix elements are also known, e.g. for a
spin chain in an anti-ferromagnetic material, the off-diagonal elements
are positive, as the case for many atomic or molecular systems.  We then
have
\begin{equation}
\label{eq:probs2}
 p_{k\ell}(t) = c_{k\ell}(t) + 2\sum_{\nu>\mu} 
      a_{k\ell;\mu\nu} e^{-t\Gamma_{\mu\nu}}\cos(\omega_{\mu\nu}t)
\end{equation}
with $a_{k\ell;\mu\nu}=s_{k\ell;\nu} s_{k\ell;\mu}$ and $c_{k\ell}=\sum_\nu
s_{k\ell;\nu}^2$, which further simplifies the reconstruction.

\section{Bayesian Parameter Estimation}

\begin{figure}
\includegraphics[width=0.5\textwidth]{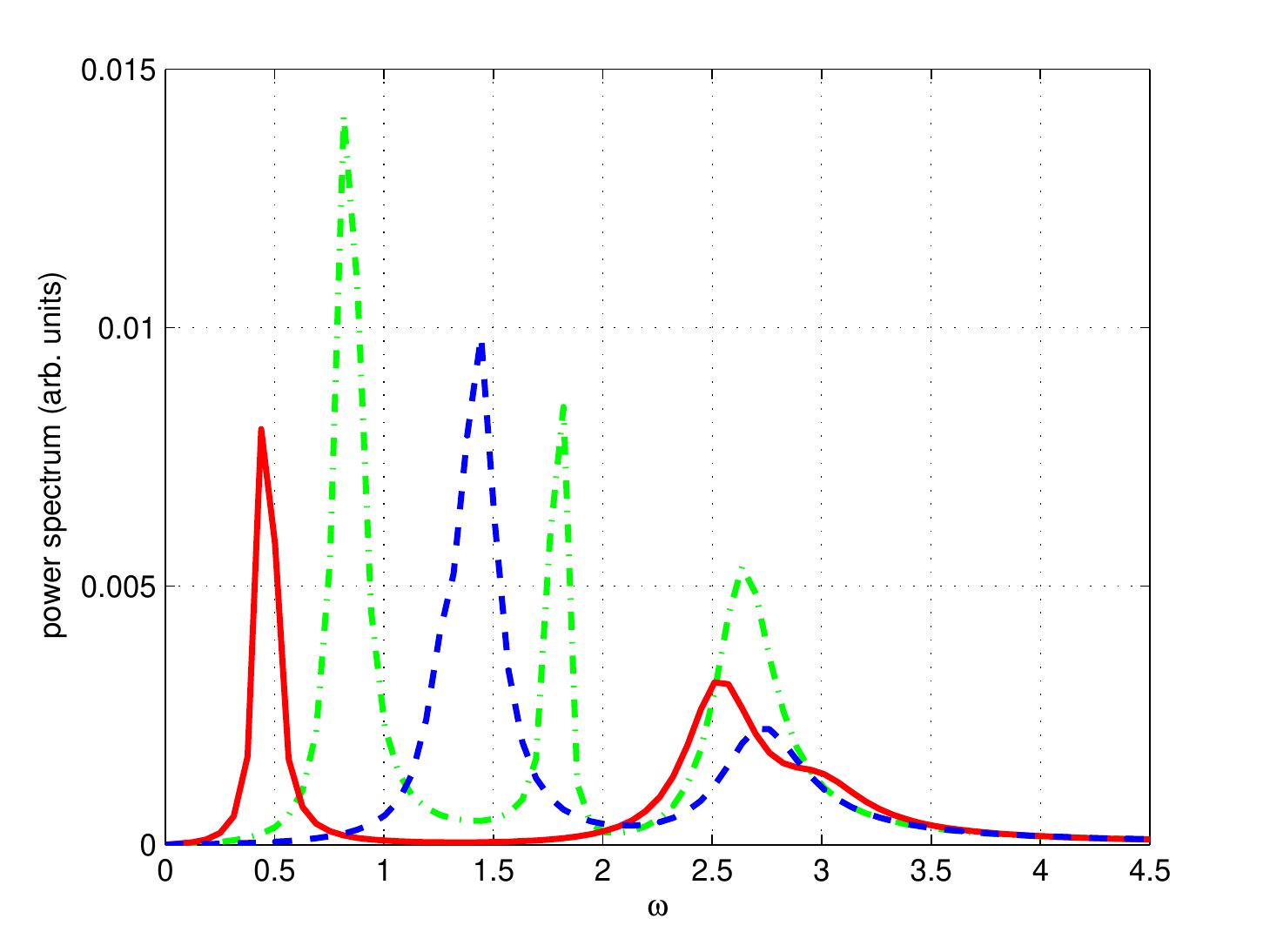}
\caption{Qutrit Power Spectra. Exponential damping broadens the peaks
leading to difficulties in accurately determining their frequencies
(green dot-dashed curve). \textit{In extremis} peaks merge (red solid
curve) and fewer than three frequencies can be observed (blue dashed
curve). However the power spectrum can be used as an initial starting
point for Bayesian estimation.}  \label{fig:Spectrum}
\end{figure}

This shows that the identification problem for dephasing that acts in
the system's natural basis is similar to the Hamiltonian identification
problem except that we also have to determine the dephasing rates
$\Gamma_{\mu\nu}$.  From the measurement results obtained from time
traces like in Fig.~\ref{fig:DataSignals}, we must extract signal
frequencies $\omega_{\mu\nu}$ and damping rates $\Gamma_{\mu\nu}$ as
well as the amplitudes $c_{k\ell}$, $a_{k\ell;\mu\nu}$ and
$b_{k\ell;\mu\nu}$ in order to be able to perform reconstruction of the
system dynamics.

We can do this again by Bayesian estimation, maximizing the likelihood
that a particular process generated the observed signal~\cite{SO2009}.
For convenience we label the transition frequencies of the system
$\omega_m$, assuming $\omega_{m+1}>\omega_m>0$, and the corresponding
dephasing rates $\Gamma_m$, and define the vectors $\gvec{\omega}
=(\omega_m)$, $\gvec{\Gamma}=(\Gamma_m)$,
$\vec{a}_{k\ell}=(a_{k\ell;m})$ and $\vec{b}_{k\ell} =(b_{k\ell;m})$
where $k,\ell$ range from $1$ to $N$ and $m$ from $1$ to the number of
transition frequencies $M$.  According to Eq.~\ref{eq:probs}, the traces
should be linear combinations
\begin{equation}
   p_{k\ell}(t_n) = 
   \sum_{m=1}^3 a_{k\ell,m} g_{2m-1}(t)+b_{k\ell,m} g_{2m}(t)+c_{k\ell} 
\end{equation}
of the $m_b = 2M+1$ basis functions
\begin{subequations}
\label{eq:case1}
\begin{align}
 g_{2m-1}(t) &=e^{-t\Gamma_m}\cos(\omega_m t),\\
 g_{2m}(t)   &=e^{-t\Gamma_m}\sin(\omega_m t),\\
 g_{2M+1}(t) &=1,
\end{align}
\end{subequations}
or in the case where $H$ is real-symmetric, the $m_b=M+1$ basis
functions
\begin{subequations}
\label{eq:case2}
\begin{align}
 g_{m}(t)   &=e^{-t\Gamma_m}\cos(\omega_m t),\\
 g_{M+1}(t) &=1,
\end{align}
\end{subequations}
and our objective is to find parameters $\Gamma_m$, $\omega_m$,
$a_{k\ell;m}$, $b_{k\ell;m}$ and $c_{k\ell}$ that maximize the
likelihood of the measured data
\begin{equation}
 L(\vec{a}_{k\ell},\vec{b}_{k\ell},c_{k\ell},\gvec{\omega},\sigma)  
  = \prod_{k,\ell=1}^N \sigma_{k\ell}^{-N} \exp\left[
  -\frac{\norm{\vec{p}_{k\ell}-\vec{d}_{k\ell}}_2^2}{2\sigma_{k\ell}^2}\right].
\end{equation}

We can eliminate the explicit dependence on the linear coefficients
$\vec{a}_{k\ell}$, $\vec{b}_{k\ell}$, $c_{k\ell}$ and the noise
variances $\sigma_{k\ell}$ by integration over suitable priors to obtain
an explicit expression for the probability of a particular model given
the observed data $\vec{d}_{k\ell}$ that depends only on the $M$ transition 
frequencies $\omega_m$ and corresponding dephasing rates $\Gamma_m$.
Following standard Bayesian analysis~\cite{88Bretthorst} we obtain
\begin{equation}
  P(\gvec{\omega},\gvec{\Gamma}|\vec{d}) \propto \prod_{k,\ell=1}^N
  \left[ 1-\frac{m_b \ave{\vec{h}_{k\ell}^2}}{N \ave{\vec{d}_{k\ell}^2}}
\right]^{(m_b-N)/2},
\end{equation}
where the averages are defined by
\begin{subequations}
\begin{align}
  \ave{\vec{d}_{k\ell}^2} &=\frac{1}{N}\sum_{n=1}^N d_{k\ell;n}^2, \\
  \ave{\vec{h}_{k\ell}^2} &=\frac{1}{m_B}\sum_{m=1}^{m_B} h_{k\ell;m}^2.
\end{align}
\end{subequations}
The components $h_{k\ell;m}$ are essentially the orthogonal projections
of the data onto a set of orthonormal basis vectors $H_m(t_n)$
\begin{equation}
  h_{k\ell;m} = \sum_{n=1}^N H_m(t_n) d_{k\ell;n}
\end{equation}
derived from the (non-orthogonal) basis functions $g_m(t)$ defined
above, evaluated at the respective sample times $t_n$, via
\begin{equation}
  H_m(t_n) = \frac{1}{\sqrt{\alpha_m}} \sum_{m'=1}^{m_B} e_{m' m} g_{m'}(t_n),
\end{equation}
where $e_{m'm}$ is a $m_b\times m_b$ matrix whose columns $\vec{e}_m$ are
the normalized eigenvectors --- $G \vec{e}_m = \alpha_m \vec{e}_m$ ---
of the $m_b \times m_b$ matrix $G=(G_{m_1 m_2})$ with
\begin{equation}
  G_{m_1 m_2} = \sum_{n=1}^N g_{m_1}(t_n) g_{m_2}(t_n).
\end{equation}
Thus, the parameter estimation problem for a system with decoherence
acting in the Hamiltonian basis is similar to that for a Hamiltonian
system, except that the sine and cosine basis functions for the Bayesian
analysis must be modified to damped sinusoids with unknown damping
rates.

The objective is to find the frequencies $\gvec{\omega}$ and damping
rates $\gvec{\Gamma}$ that maximize $P(\gvec{\omega},\vec{\Gamma}|
\vec{d}_{k\ell})$, or equivalently, the log-likelihood function
\begin{equation}
 \label{eq:logP}
  \log_{10} P(\gvec{\omega},\gvec{\Gamma}|\vec{d}_{k\ell}) 
  = \frac{m_b-N}{2} \sum_{k,\ell=1}^N  \log_{10} 
   \left[ 1 - \frac{m_b \ave{\vec{h}_{k\ell}^2}}{N \ave{\vec{d}_{k\ell}^2}} \right].
\end{equation}
Given a solution $\gvec{\omega}$ and $\gvec{\Gamma}$ that maximizes this
log-likelihood, it can be shown that the corresponding optimal
coefficients in the general case (\ref{eq:case1}) are
\begin{subequations}
  \begin{align}
    \vec{a}_{k\ell} &=
    \left(\ave{x_{k\ell;1}},\ave{x_{k\ell;3}},\ldots,\ave{x_{k\ell;m_B-2}}
    \right), \\
    \vec{b}_{k\ell} &=
    \left(\ave{x_{k\ell;2}},\ave{x_{k\ell;4}},\ldots,\ave{x_{k\ell;m_B-1}}
    \right), \\
    c_{k\ell} &= \ave{x_{k\ell;m_B}},
\end{align}
\end{subequations}
where $\ave{x_{k\ell;m}}$ is shorthand notation for the expectation
values $E(x_{k\ell;m}|\gvec{\omega},\gvec{\Gamma},\vec{d}_{k\ell})$ of
the linear coefficients of the basis functions, given the optimal
frequencies $\gvec{\omega}$ and damping rates $\vec{\Gamma}$ and the
data $\vec{d}_{k\ell}$.  Similarly in the special case~(\ref{eq:case2})
\begin{subequations}
\begin{align}
  \vec{a}_{k\ell} &=
  \left(\ave{x_{k\ell;1}},\ave{x_{k\ell;3}},\ldots,\ave{x_{k\ell;m_B-1}}\right),\\
  c_{k\ell}       &= \ave{x_{k\ell;m_B}}.
\end{align}
\end{subequations}

Since the log-likelihood function is sharply peaked with generally many
local extrema, finding the global optimum using gradient-type
optimization algorithms starting with a completely random guess for
$\gvec{\omega}$ and $\gvec{\Gamma}$ is inefficient.  A global
optimization such as pattern search or evolutionary algorithms might
circumvent this problem, but neither proved either very effective in our
case, especially for higher-dimensional search spaces.  Alternatively,
starting with a somewhat reasonable initial guess, especially for the
frequencies, a standard quasi-Newton optimization method with cubic line
search~\cite{BFGS1,BFGS2,BFGS3,BFGS4} proved generally very effective in
finding the global maximum.

To obtain an initial estimate for the frequencies we used the sum of the
power spectra of the signals.  Although the peaks in the power spectrum
are not optimal frequency estimators when there are multiple frequencies
and the exact peak locations can be difficult to ascertain even for
systems with only three frequencies, as Fig.~\ref{fig:Spectrum} shows,
rough estimates of the peak locations usually seem to provide a
reasonable initial guess for the gradient-based likelihood optimization
routine. In principle the damping rates could be estimated from the
peaks widths as well but these estimates can be tricky, especially for
overlapping and minor peaks, hence we chose multiple runs with random
initial guesses for the damping rates $\gvec{\Gamma}$ and selected the
run with the highest final likelihood (``global'' maximum).

Given the extracted signal parameters we have to solve two further
inverse problems: (i) reconstructing the level structure from the
frequencies and (ii) constructing the matrix $W$ that relates the
Hamiltonian basis to the computational basis.  The former usually
involves analyzing the relationships between the frequencies as
illustrated in~\cite{SO2009}.  In general this is be tricky but for a
qutrit system, is analysis is essentially trivial.  The basis
reconstruction requires solving further optimization problems to find
the coefficients $s_{k\ell;\mu}$ such that Eqs~(\ref{eq:coeff}) are
satisfied given the estimates for the parameters $a_{k\ell;\mu}$,
$b_{k\ell;\mu}$, $c_{k\ell}$ and $\Delta_{k\ell;\mu\nu}$ derived in the
previous step.  Due to finite sampling and noise, the inversion may not
be exact, hence we recast it as a constrained optimization problem and
solve it as described in~\cite{SO2009}.

Our previous analysis~\cite{SO2009} also shows that we can only identify
a single generic Hamiltonian up to equivalence
\begin{equation}
  H \simeq D^\dag \tilde{H} D + \lambda \ONE,
\end{equation}
where $D=\diag(1,e^{i\delta_{12}},\ldots,e^{i\delta_{1N}})$ is a
diagonal unitary matrix, in the basis of the measurement. However, if
the off-diagonal elements in the Hamiltonian are known to be real and
positive, for instance, then the Hamiltonian will be uniquely determined
up to a global energy level shift $\lambda\ONE$, at least in the generic
case, since we have $|H_{k\ell}|=|\tilde{H}_{k\ell}|$ for $k\neq\ell$.
For a quantum control situation, the system dynamics can be controlled
and hence different Hamiltonians can be applied, and in the case
subsequent Hamiltonians can be fully determined up to the gauge fixed by
the initial Hamiltonian.  By varying control parameters and tracking the
change in the system dynamics, a dynamical control model can be built of
the system.

\section{Results}

We randomly generated 100 real-symmetric qutrit Hamiltonians and dephasing
operators with different spectral properties and the geometric average of the
system $Q$-factors ranging from 12 to 72. From these we generated various data
traces corresponding to the stroboscopic sampling described in section
\ref{sect:exp}.  We considered three cases, the zero noise case
($N_\infty=$infinite samples per point), fixed finite sampling with
$N_e=N_{1000}=1000$ experimental repetitions per time point, and an adaptive
sampling strategy $N_{var}$ which varies the number of samples per point to
reach an estimated target signal to noise ratio of $\ave{p_{k\ell}(t)} \ge
10/\sqrt{N_e}$ for all $k,\ell$ and $t$ with an upper limit of $N_e\le 10,000$
for each data point.  We then applied our parameter estimation and
reconstruction algorithms to the resulting data traces.  A range of dephasing
rates was studied to see the effect on the reconstruction of the Hamiltonian
part of the dynamics.  For the purposes of control, accurate determination of
the Hamiltonian is much more important than a precise determination of the
dephasing rate, usually it suffices to know that they are below certain limits.

\begin{table}
\begin{tabular}{| l | c | c | c | c | c | r | }
  \hline			
 &  $N_\infty$ & $N_\infty^H$ & $N_{1000}$ & $N_{1000}^H$ & $N_{var}$ &
$N_{var}^{H}$  \\\hline
$\bar{L}$
& 5.9e04& 4.9e04& 1.2e04& 1.3e04& 1.4e04& 1.6e04\\ \hline
$\bar{\epsilon}_\omega$
& 1.8e-07& 2.8e-07& 6.6e-04& 2.5e-05& 4.9e-04& 1.6e-05\\ \hline
$\bar{\epsilon}_\Gamma$
& 7.2e-06&        & 1.6e-02&        & 1.2e-02&       \\ \hline
$\bar{\epsilon}_a$
& 1.6e-05& 4.3e-06& 4.4e-01& 6.4e-02& 1.9e-01& 3.1e-02\\ \hline
$\bar{\epsilon}_S$
& 2.6e-06& 4.4e-07& 2.3e-02& 2.7e-03& 1.3e-02& 2.0e-03\\ \hline
$\bar{\epsilon}_H$
& 3.7e-06& 1.2e-06& 1.8e-02& 2.5e-03& 1.3e-02& 1.9e-03\\ \hline
\end{tabular}
\caption{Median Likelihoods ($\bar{L}$) and Error Rates ($\bar\epsilon$) for Qutrit
  Systems. For the 100 qutrit systems we compared the case with and without
  dephasing (superscript H) for different samples ($N_e$) per
  data point.  With no sampling noise $N_\infty$, there was a
  small change in the median errors. For the $N_{1000}$ case, the
  median errors  increase due to the  sampling noise, the addition of dephasing
  increases the final error by an order of magnitude to the $1\%$ region.
  A simple adaptive scheme $N_{var}$ does similarly. The Hamiltonian is
  reconstructed using several runs of the optimization routine, and the
  solution with the minimum basis error is chosen.}
\label{tab:median}
\end{table}

Table~\ref{tab:median} shows the median errors for various cases. Comparing the
dephasing/no dephasing cases, the errors are similar in the absence of
projection noise $(N_\infty$). The frequency $\omega$ estimation is slightly
more accurate but estimation of the signal amplitudes $a_{k\ell;\mu\nu}$ is
slightly less accurate since the basis functions depend on $\Gamma$, hence
errors in both $\omega$ and $\Gamma$ contribute to errors in the coefficients
$a_{k\ell;\mu\nu}$.  For reduced signal to noise, dephasing decreases the
maximum likelihood and increases frequency, basis and reconstructed Hamiltonian
errors with a marked increase in median of the amplitude errors.  Adaptive
sampling overall increases the accuracy of the parameter estimation step and the
reconstructed Hamiltonian for both Hamiltonian and dephasing systems but the
improvement is more pronounced for dephasing systems.  This may be due to
adaptive sampling being more beneficial for small signal amplitudes i.e.,
decaying signals.  This suggests the use of adaptive sampling to increase the
signal to noise ratio for samples at increasing times. Alternatively, the sample
data can be weighted to give precedence to earlier samples. Further exploration
of these methods will be the subject of future study.

Dephasing leads to a reduction in signal at long times which can lead us
to fitting noise. For strong dephasing, this leads to reduced accuracy
in the estimation of the frequencies, and hence increased errors in the
other parameters.  The spread in the Fourier peaks can also lead to
problems for closely spaced frequencies.  This in itself is not a
problem \textit{per se} for the Bayesian parameter estimation
step~\cite{SO2009}, except that it can lead to inaccurate initial search
parameters coming from peak detection in the power spectrum. This can be
obviated somewhat by trying different initial parameters assuming that
either of the two remaining peaks were doublets and using the most
likely result.

For systems of interest for quantum information processing, the
dephasing rates should be sufficiently low so that the damping of the
Rabi-type oscillations do not impact the scheme greatly. For very small
dephasing rates, However, it can be a problem if the algorithm
overestimates the dephasing rates which means that the basis functions
used are not suitable, and this is reflected in errors of the estimated
amplitudes. For such systems, it is a simple enough matter to test
models which are purely Hamiltonian to see which gives the larger
likelihood.

\begin{figure}
\includegraphics[width=0.5\textwidth]{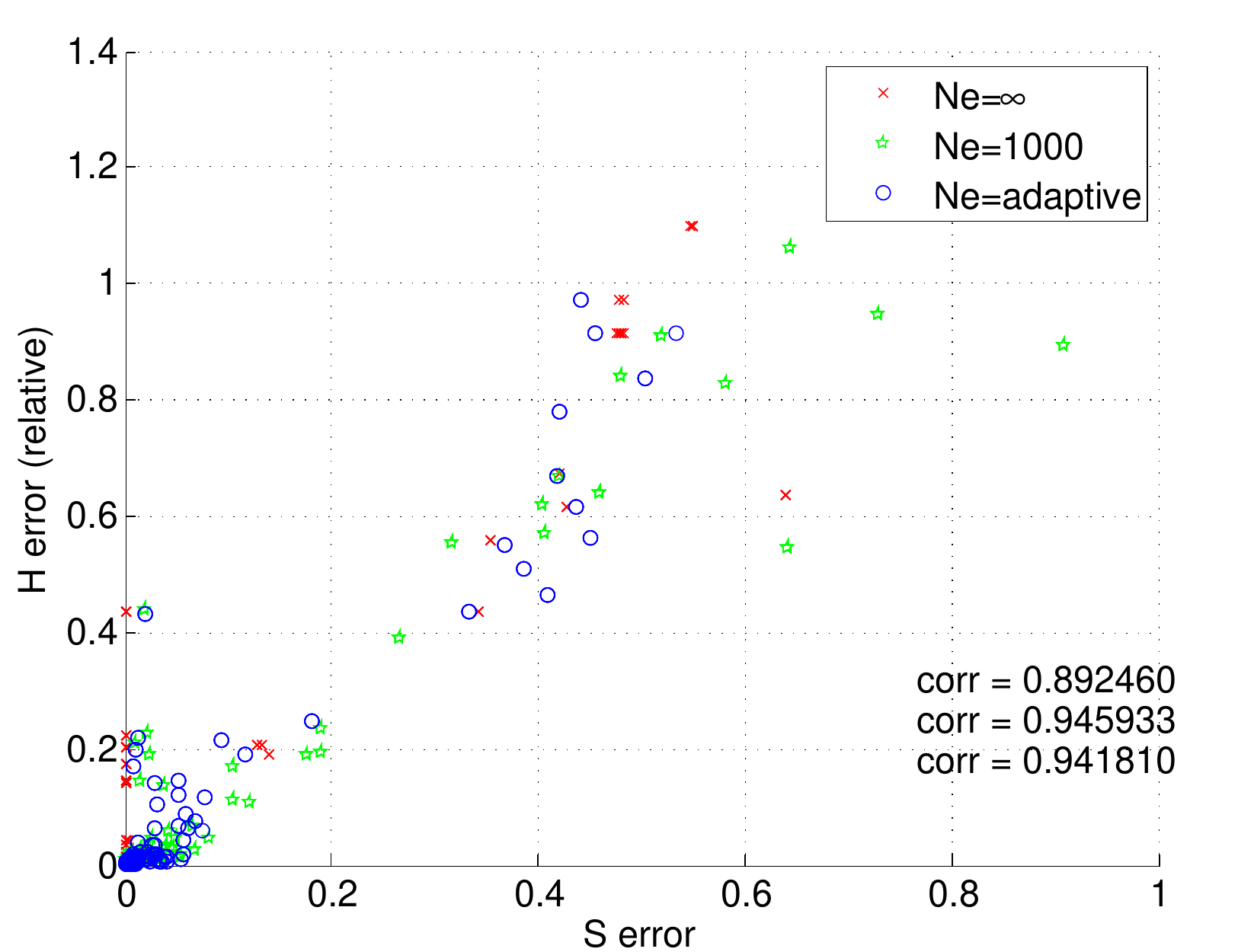}
\caption{Hamiltonian vs Basis Reconstruction Error $S$ for various
samplings shows a strong correlation and suggests that the total error
in the Hamiltonian is dominated by errors in the basis reconstruction
step that comes about from the separate optimization of each basis
function from the amplitude estimation, which may not lead to orthogonal
data vectors.}
\label{fig:SError}
\end{figure}

One factor which limits the reconstruction is that we may obtain a set
of $N\times N$ matrices $P_\nu=(s_{k\ell;\nu})$, which ideally should 
be projectors onto orthogonal eigenspaces, but may not always form an
orthogonal set of projectors.  We can quantify this basis error by
\begin{equation}
  S = \max_{\mu\nu} \left|\Tr(P_\nu^\dag P_\mu) - \delta_{\mu\nu} \right|.
\end{equation}
Fig.~\ref{fig:SError} shows that there is a strong correlation between
$S$ and the (relative) error in the final reconstructed Hamiltonian.
Thus, we can use $S$ to choose the best reconstructed Hamiltonian from
multiple optimization runs and as a rough indication of the likely
accuracy of the reconstructed Hamiltonian.  The data also suggests that
there is little direct correlation between the likelihood and errors in
the parameter estimation step and the final Hamiltonian error,
suggesting that the final error in the Hamiltonian is dominated by
errors in the basis reconstruction step.  The reconstruction step
obviously depends on the parameter estimates obtained in the first step,
and poor estimates for the parameters will generally result in large
Hamiltonian errors, but in some cases the basis reconstruction produces
poor results even when the individual errors in the estimated parameters
are small.  It should be possible to improve the reconstruction step by
solving the $N^2$ optimization problems for the $s_{k\ell;\mu}$
simultaneously rather than independently and enforcing orthonormality
constraints for the basis vectors, but doing so would require solving a
rather more complicated optimization problem with several nontrivial
constraints.

\section{Discussion}

Other researchers have also begun to address the problem of system
characterization with limited resources.  For example, Leghtas et
al.~\cite{LMR2009} also consider estimating parameters of three-level
quantum systems using weak continuous population measurements. However,
in their case it is assumed that most of the system is already known
including the transition frequencies and the precise structure of the
Hamiltonian, and there is no intrinsic decoherence. They consider
extracting only two real parameters of the system, the dipole transition
strengths between levels 1-2 and 2-3, which simplifies the problem
enormously.

Burgarth et al.~\cite{BMN2009,BK2009} also consider Hamiltonian
characterization with restricted resources for Heisenberg spin chains
where only a small subset of spins are individually addressable.  The
form and structure of the Hamiltonian is known \textit{a priori} to be
of a particular class, and only the coupling strengths and anisotropy of
the system Hamiltonian are to be determined. The sign of the couplings
is also known beforehand. Characterization is achieved in this case by
preparing different initial states of the first spin, letting the system
evolve and then performing quantum state tomography on the accessible
spins. If we consider a system of three spins, the first excitation
subspace acts as a qutrit. Our protocol could be applied to this problem
with some modifications. Our scheme does not require state tomography,
only the determination of position of the up-spin, and there is no
requirement to know the network topology. It would be interesting to
explore Bayesian analysis of the response of such systems for
Hamiltonian characterization, and especially the role of topology in
identifiability, and whether it is possible to relax the requirement for
addressability of all spins.

In summary, we have shown that our current two-step procedure of
Bayesian parameter estimation followed by a reconstruction via
optimization works in the presence of dephasing on three-level
systems.  However, we find that the reconstruction step is a weak point
of our current implementation.  It may be possible to eliminate the
parameter estimation step and directly apply Bayesian maximum likelihood
estimation upon the dynamical system parameters. This would have the
advantage of always giving admissible solutions at all steps. Another
direction which should be explored is adaptive sampling, not only
varying experimental repetitions per data point, but also using
non-uniform time-domain sampling for better frequency discrimination.

\end{document}